\documentclass[journal,letters]{IEEEtran}

\usepackage[final]{graphicx}
\graphicspath{ {./} }
\usepackage[reqno]{amsmath}
\usepackage{amssymb}
\usepackage{amsthm}
\usepackage{subfig}
\usepackage{epstopdf}
\usepackage{algorithm}
\usepackage{algorithmicx}
\usepackage{algpseudocode}
\usepackage{setspace}
\usepackage[T1]{fontenc}
\usepackage{color,soul}
\usepackage{multirow}
\usepackage{array}
\usepackage{cite}

\usepackage{verbatim}

\usepackage{enumitem}
\usepackage{url}
\usepackage{csvsimple}
\usepackage{float}

\usepackage{float}
\usepackage{graphicx}
\usepackage{caption}

\usepackage{geometry}
\usepackage{array}
\newcommand{\PreserveBackslash}[1]{\let\temp=\\#1\let\\=\temp}
\newcolumntype{C}[1]{>{\PreserveBackslash\centering}p{#1}}
\newcolumntype{R}[1]{>{\PreserveBackslash\raggedleft}p{#1}}
\newcolumntype{L}[1]{>{\PreserveBackslash\raggedright}p{#1}}

\begin{document}

\title{A Study on Transferability of Deep Learning Models for Network Intrusion Detection}

\author{\large{Shreya Ghosh$^{\dagger}$, {\em Student Member, IEEE}}, \large{Abu Shafin Mohammad Mahdee Jameel$^{\dagger}$, {\em Student Member, IEEE}}, \large{Aly El Gamal$^{\dagger}$, {\em Senior Member, IEEE}\vspace{-10pt}}
\thanks{$^{\dagger}$ The authors are with the School of Electrical and Computer Engineering of Purdue University. (e-mail: \{ghosh64, amahdeej, elgamala\}@purdue.edu)}
\vspace{-10pt}}

\maketitle

\begin{abstract}

In this paper, we explore transferability in learning between different attack classes in a network intrusion detection setup. We evaluate transferability of attack classes by training a deep learning model with a specific attack class and testing it on a separate attack class. We observe the effects of real and synthetically generated data augmentation techniques on transferability. We investigate the nature of observed transferability relationships, which can be either symmetric or asymmetric. We also examine explainability of the transferability relationships using the recursive feature elimination algorithm. We study data preprocessing techniques to boost model performance. The code for this work
can be found at https://github.com/ghosh64/transferability.

\end{abstract}
\IEEEpeerreviewmaketitle
\section{Introduction}

\IEEEPARstart{I}{n} transfer learning, a machine learning model uses knowledge learnt from past trainings to improve generalization on similar tasks. Many computer networks employ machine learning models to detect malicious data packets entering a network. In this paper, we explore using transfer learning to aid machine learning based intrusion detection softwares to detect novel or newer attacks. 
Our goal is to leverage learning transferability from select attacks in the training dataset that causes the model to generalize for other unknown attacks which can be extended to rare attacks. 

It is difficult to provide guarantees about the kind of attacks that a computer network can expect to see and is susceptible to, especially with the rise in the number of zero day attacks. Current machine learning algorithms are usually trained to detect a set of known attacks, or learn from attacks performed in the past. Model learning, if transferable, would enable these models to be able to detect a wider range of attacks than just the attacks they have been trained to detect. Without transferability studies, it is hard to predict the true range of attacks for which the employed machine learning algorithm is actually robust.

For groups of similar attacks, transfer learning enables us to find representative attacks from that group that help us train the model with a subset of attacks yet achieve the same level of security as being trained with the full set of attacks in the group. Increasing applications of machine learning has also seen deep learning models going into hardware chips. Training a model on a subset of attacks decreases training time, is computationally efficient and has a lower memory requirement, while performing at par with a model trained on a full training set. This is important especially if this algorithm is deployed on a resource constrained device.

Research in this domain has led to the availability of standardized network traffic datasets. Some of the common datasets used are CAIDA 2007\cite{caida2007}, DARPA 98, KDD 99 \cite{tavallaee2009detailed}, to name a few. We use the widely adopted traffic dataset CICIDS 2017\cite{sharafaldin2018toward} for training our model. 
These datasets provide a complete description of network packets, but they have a high dimensional feature space, making it non-trivial for machine learning models to learn meaningful information. 

Classic approaches on network detection depended on different approaches like Naive-Bayes classifiers \cite{koc2012network}, Random Forest classifiers \cite{zhang2008random}, and Support Vector Machines \cite{chang2017network}. However, multiple recent studies have shown that deep learning based methods provide superior performance compared to traditional approaches \cite{alkasassbeh2016detecting, lopez2019network, vinayakumar2019deep}. Therefore, a lot of the recent studies in this domain focus on deep learning algorithms \cite{barnard2022robust, otoum2019feasibility}, and these approaches are taken as benchmark during dataset developments \cite{koroniotis2019towards}.

Most current day research focuses on developing deep learning architectures that are able to extract important information from large datasets. Although these models are highly accurate in detecting known attacks, they are not always robust against novel attacks. In addition, anomaly detection based studies often treat all unknown network intrusions as a single class and as such do not provide a good analysis of the transferability of existing attacks on separate unique attack classes. This leaves transferability for rare attack detection as an important new development in this area of research. Due to this, some of the most recent works in this domain focus on transferability \cite{verkerken2022towards, catillo2022transferability}. Transferability study of a model is concerned with training a deep learning model using data from one source and deploying it to detect data from another source. In \cite{verkerken2022towards}, the authors train their models on one dataset and test it on a separate dataset, with focus on one attack class. The study in \cite{verkerken2022towards} improves on this by focusing on transferability between attack classes, i.e. when the model is trained on one attack class and tested on another. However, the authors pre-select training and testing attack classes. 

In this paper, we present a comprehensive transferability study by designing our experiment to avoid dividing the attack classes between training and testing, which enables us to examine and comment on the symmetry or asymmetry in transferability relationships. We also study the effect of input features on the transferability by inspecting feature selection methods. We examine the effects of data augmentation using either bootstrapped real data or using synthetically generated data. Finally, we examine three different preprocessing techniques to identify a technique that boosts model performance. In this regard, the contributions of this paper are as follows:

\begin{itemize}
    \item We develop an architecture that is able to extract meaningful data from high dimensional training data and limited number of attack data packets.
    \item We analyse transferability relationships by examining the effects of training and testing on separate attack classes, and observe underlying reasons for symmetric or asymmetric transferability relationships in training-testing attack pairs.
    \item We test the effects of training data augmentation methods using bootstrapped real data or synthetically generated attack data.
    \item Using an analysis of the most important features between attack classes, we establish explainability of transferability relationships based on feature selection.
    \item We implement data preprocessing techniques that can boost transferability.
\end{itemize}
This study will be useful in training design of deep learning based intrusion detection systems by providing a more nuanced understanding of the training transferability of various common intrusion types.

\begin{table}[t]
\centering
\captionsetup{justification=centering}
\caption{Dataset Composition}
\label{tab:table-1}
    \begin{tabular}{| R{1.0cm} | L{3.7cm} | L{1.6cm} |}
    \hline
    \textbf{Classes} & \textbf{Attack}          & \textbf{Percentage of data} \\ \hline
    0                & BENIGN                   & 80.3\%                        \\ \hline
    1                & Bot                      & 0.069\%                       \\ \hline
    2                & DDoS                     & 4.52\%                        \\ \hline
    3                & DoS GoldenEye            & 0.36\%                        \\ \hline
    4                & DoS Hulk                 & 8.16\%                        \\ \hline
    5                & DoS Slowhttptest         & 0.19\%                        \\ \hline
    6                & DoS Slowloris            & 0.2\%                         \\ \hline
    7                & FTP-Patator              & 0.28\%                        \\ \hline
    8                & Heartbleed               & 0.00038\%                     \\ \hline
    9                & Infiltration             & 0.0012\%                      \\ \hline
    10               & PortScan                 & 5.61\%                        \\ \hline
    11               & SSH-Patator              & 0.2\%                         \\ \hline
    12               & Web Attack Brute Force   & 0.053 \%                     \\ \hline
    13               & Web Attack Sql Injection & 0.00074 \%                   \\ \hline
    14               & Web Attack XSS           & 0.023 \%                     \\ \hline
    \end{tabular}
\end{table}

\begin{figure*}[h]
    \centering
    \includegraphics[width=1.0\textwidth]{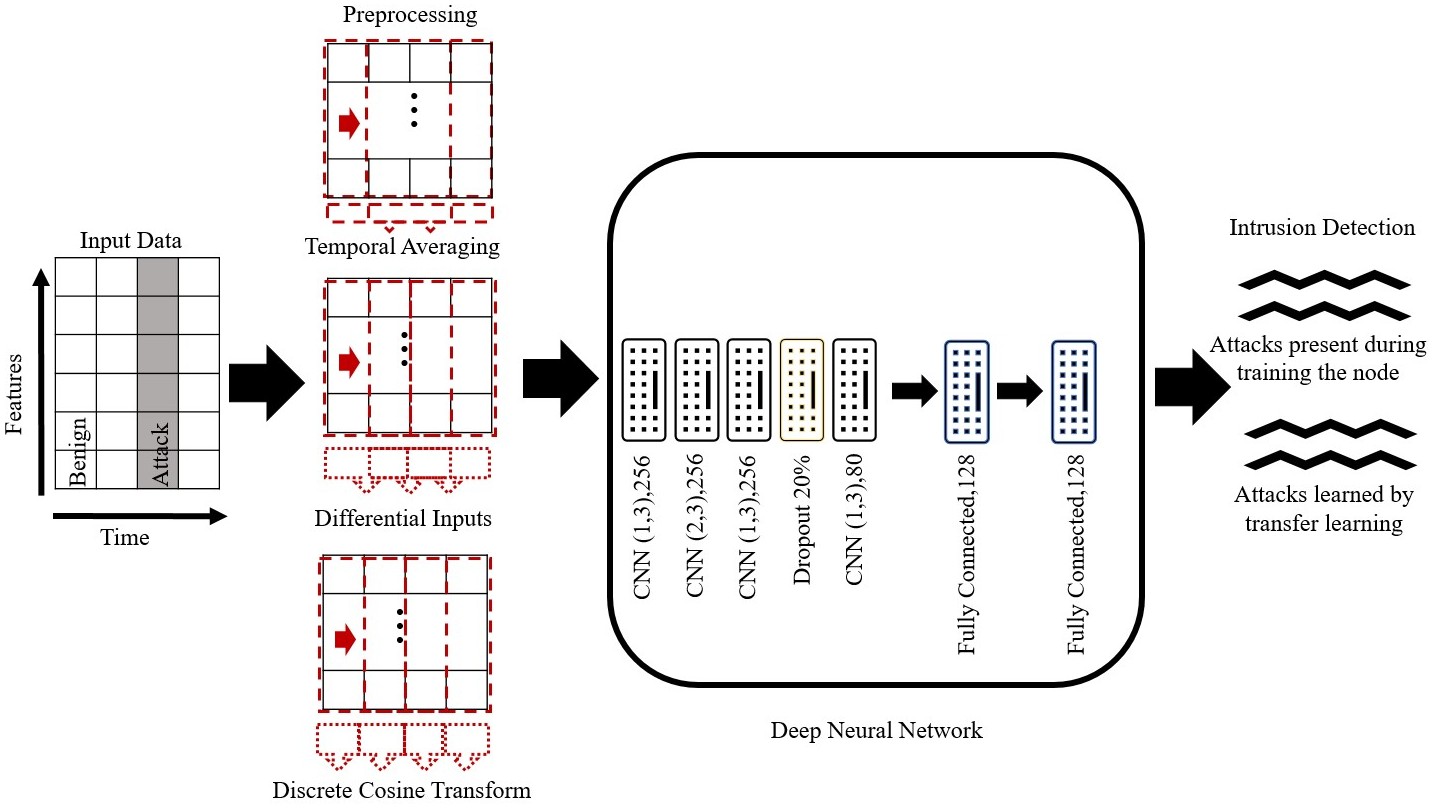}
    \caption{Architecture of Proposed Intrusion Detection Model \vspace{-15pt}}
    \label{fig:network}
\end{figure*}

\section{Dataset}

The dataset used to train our model is generated by the Canadian Institute of Cybersecurity: CIC-IDS 2017 \cite{sharafaldin2018toward}. This dataset consists of 78 features and 14 types of attacks.the dataset has class imbalance - while 80\% of the data is benign, only 20\% of the dataset is attack data.  The composition is further elaborated in Table \ref{tab:table-1}. 

\section{Methodology}

To develop our proposed architecture, we train the model with data from all classes and test the model on a testing set that has a similar distribution of data. The under represented attack classes are not augmented during training to determine which model is able to perform deep feature extraction even with a low number of training samples for certain classes. To address the issue of severe under-representation of several classes in the CICIDS 2017 dataset, we evaluate two techniques during transferability analysis to increase the representation of each attack class:

\subsubsection{SMOTE generated data}

We generate synthetic attack data for each attack class using Synthetic Minority Oversampling Technique (SMOTE) \cite{smote}. Using this technique (with the parameter $k$ set at $5$), we identify specific regions in the feature space that can be used to generate more samples belonging to a certain class of data. The number of attack packets generated using this technique is used to match the number of benign data packets in the dataset. This process is repeated for each attack.

\subsubsection{Bootstrapped dataset} 

Another technique used to create a balanced training dataset is for every attack that the model is trained with, the original attack data is resampled to match the number of benign data packets. This process is used to replicate attack data corresponding to all attack classes.

While developing our proposed architecture, we first train the model with real data to identify the model that can best classify the real training data even with low representation in some classes. Then, during our transferability study, the proposed architecture is trained with both types of enhanced datasets and tested on real attack data only. We create the training dataset as 50\% benign data and 50\% attack data. We repeat this for training the model with all attacks in the dataset. 

\subsection{Proposed Architecture}

We developed the proposed architecture through an empirical study maximizing attack data classification accuracy for the multi-class classification problem. This architecture is able to perform deep feature extraction using flow based features not only for adequately sampled attacks, but also for severely under represented attacks. All 78 features from the dataset were used for training. The model is trained on 60\% of data and tested on 30\% of the data. 10\% of the training data is used as validation data. The model is trained for 100 epochs.

\subsection{Analysis of Transferability}

Using feature analysis, we make inferences about the observed symmetric or asymmetric transferability. To identify important features, we use the Recursive Feature Elimination (RFE) algorithm. The CICIDS 2017 dataset has 78 features and our goal here is to identify a smaller subset of features that are important for learning and generalizability of our model. Having a smaller feature set would also speed up training and inference times. 

We utilize two implementations of the RFE algorithm. For the first implementation, we select a single attack and use the RFE algorithm to identify features that the classifier uses to separate attack data from the benign data. We repeat this process for all attacks in the dataset and identify the common features between two attack classes that can allow a model to be transferable between attack classes. 

The second implementation involves taking data from two attack classes along with benign data. Then we use the RFE algorithm to identify a reduced feature set that separates attack data from benign data. For example, two correlated attacks DoS GoldenEye and DoS Hulk are both labelled as attack data and the RFE algorithm chooses features that are important for the model to differentiate benign data from attack data corresponding to both classes. We posit that the features selected using such an algorithm are important to both the attack classes. In a test setting where our goal is to transfer a model trained on an attack class to work on another attack class, presence of highly correlated features in this experiment can give us an indication of transferability. 

\subsection{Data Preprocessing} 

We propose to utilize data preprocessing steps that can boost transferability. In addition, some data preprocessing techniques can help preserve the privacy of the data, as the deep learning model is trained with the preprocessed data and it never sees the actual data. We study three different preprocessing steps, namely differential inputs, temporal averaging and discrete cosine transforms as preprocessing steps to boost the transferability.

\subsubsection{Differential Inputs}

Each input sample to the model is the difference of the current input sample and the previous input sample. Mathematically this can be seen as:

\begin{center}
    $y_{t}=x_{t}-x_{t-1}$
\end{center}
Where $x_{t}$ is the original sample for time $t$, and $y_{t}$ is the transformed sample.

\subsubsection{Temporal Averaging}

Here, each input sample to the model is the temporal average of $n$ other samples. Mathematically this can be given as:

\begin{center}

$y_{t}=\frac{\sum_{i=0}^{n-1} x(t-i)}{n}, n=\text{window size}$ 
    
\end{center}

\subsubsection{Discrete Cosine Transform(DCT)}

The DCT transform transfers the data to its frequency components using real coefficients. As Fourier transform introduces complex coefficients, which require a change in deep learning model structure, we implemented the DCT transform as a preprocessing step to investigate if it improves transferability. Mathematically this transform is given as:

\begin{center}
$y_t=x_0+(-1)^{t}x_{N-1}+2\sum_{n=1}^{N-2}x_n cos \frac{\pi tn}{N-1}$    
\end{center}
Where $N$ is the length of the DCT transform. We transform each input batch to the deep learning model separately.

\section{Results}

\begin{figure}[t]
    \centering
    \includegraphics[width=0.45\textwidth]{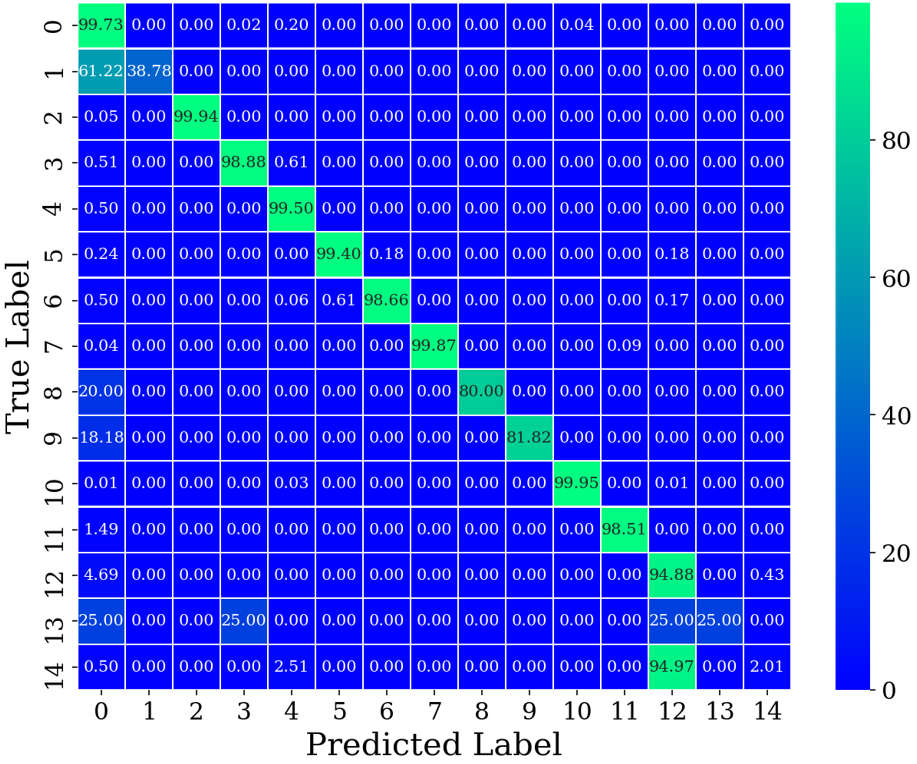}
    \caption{Confusion matrix for 15 class classification problem expressed as percentages}
    
    \label{fig:multiclass}
\end{figure}

\begin{figure}[t]
    \centering
    \includegraphics[width=0.45\textwidth]{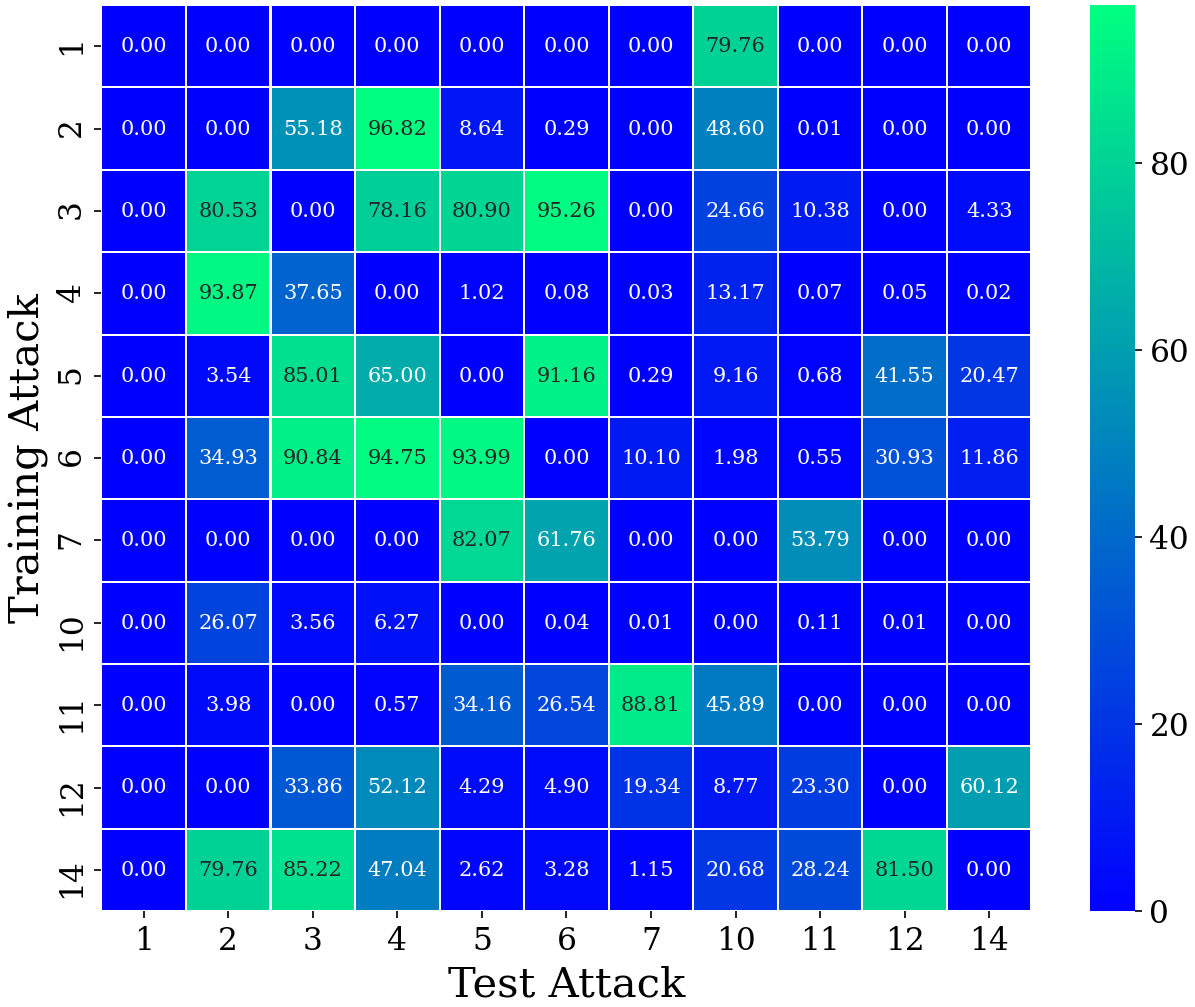}
    \caption{Attack Accuracy for different test train pairs for real training data\vspace{0pt}}
    
    \label{fig:realattackaccr}
\end{figure}

\begin{figure}[t]
    \centering
    \includegraphics[width=0.45\textwidth]{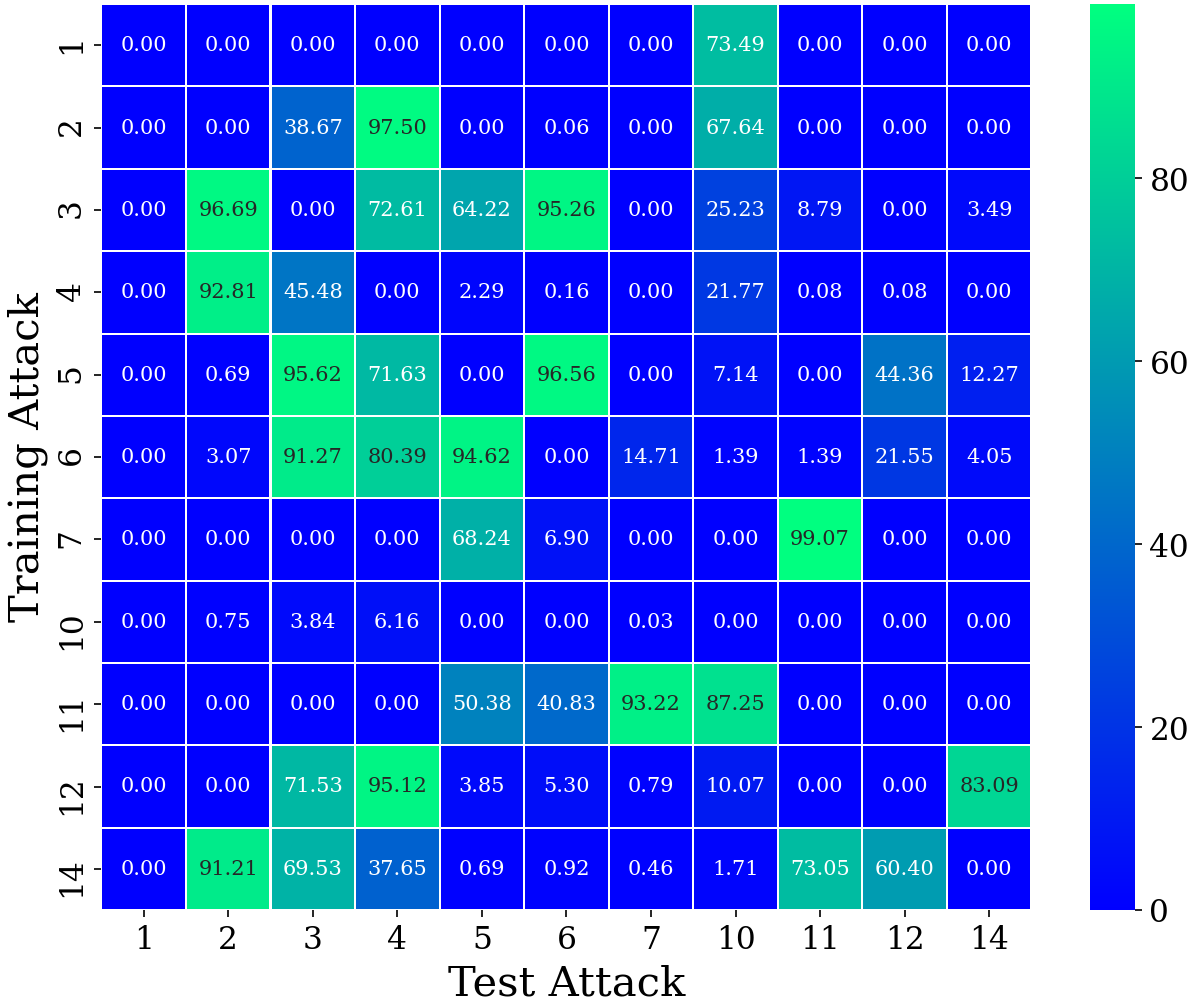}
    \caption{Attack Accuracy for different test train pairs for bootstrapped training data and real testing data}
    
    \label{fig:bootstrapaccr}
\end{figure}

\begin{figure}[t]
    \centering
    \includegraphics[width=0.45\textwidth]{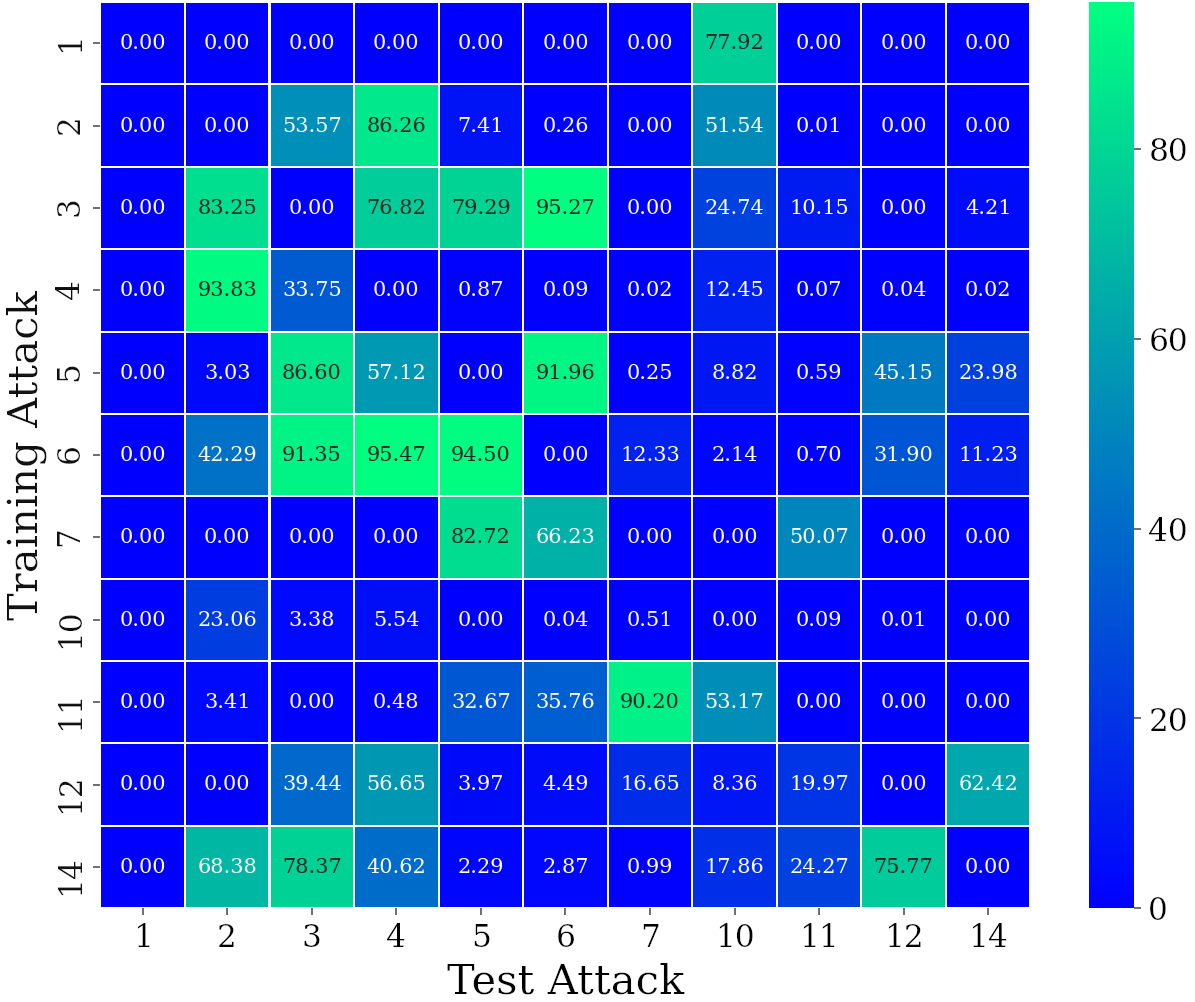}
    \caption{Attack Accuracy for different test train pairs for synthetic SMOTE training data and real testing data \vspace{-10pt}}
    
    \label{fig:synattackaccr}
    
\end{figure}

In this section, we first evaluate the performance of the proposed architecture in a multi-class scenario. Then we utilize a two-class version of the same architecture to test transferability of attacks. We investigate the effects of feature selection on transferability, which can be used to infer transferability between different attacks. Next, we explore ways to address the inherent imbalance between benign and attack data. We test two approaches, bootstrapping, and synthetic data generation. We then analyze symmetry in transferability relationships, and postulate explanations for the observed symmetry or asymmetry. Finally, we test the effects of different preprocessing techniques on transferability.

\subsection{Performance of Proposed Architecture}

First, we test the efficiency of the proposed architecture in a multi-class setup, where the network is tasked to identify individual attack classes. The confusion matrix resulting from this experiment is presented in Fig. \ref{fig:multiclass}. We can see that the multi-class classifier achieves a classification accuracy higher than 98\% for 8 out of 11 attack classes, and in case of others there is scope for improvement.
It should be noted that attacks 8, 9 and 13 are not considered for transferability analysis due to the very small number of samples available from these attacks (Attack 8 (HeartBleed) and Attack 9 (Infiltration) have 11 and 36 data packets in the dataset, respectively). Here the diagonal elements represent the percentage of data in each class that is correctly classified.

\subsection{Attack Transferablility}
Next, we focus on the study of attack transferability, where training data for a specific attack might not be available beforehand. Here, the goal is to train with a specific class and observe if the trained model performs well for any other class. For this, we use a version of the deep learning model that performs two-class classification. The model shows good classification performance, as seen from Fig. \ref{fig:multiclass}. This makes it a good fit for the attack transferability study. 

We study attack correlations by training the proposed architecture with the original unbalanced dataset, SMOTE generated dataset and bootstrapped dataset and test it for generalizability. The results for these three cases are presented in Figs. \ref{fig:realattackaccr}, \ref{fig:synattackaccr}, and \ref{fig:bootstrapaccr}, respectively. The diagonal elements in the confusion matrices are placeholders as they represent the accuracies when training and testing on the same attack that can be observed from Fig. \ref{fig:multiclass}. In Table \ref{tab:table-2}, we summarize the transferablity relationships between different attack classes based on Fig. \ref{fig:realattackaccr}.

\subsection{Bootstrapping and Synthetic Data}

We can observe from Fig. \ref{fig:bootstrapaccr} and Fig. \ref{fig:synattackaccr} that most of the training-testing attack pairs that showed transferability relationships when the model was trained with the original training data also show correlations when the model was trained with synthetic and bootstrapped data. Overall,  training the model with bootstrapped data improves the accuracy of detection for most training-testing attack pairs compared to no bootstrapping. For example, for the ordered attack pairs (3,2), (2,4), (5,3), (5,6) and (12,14), the performance is significantly boosted as seen in Fig \ref{fig:bootstrapaccr}. On the other hand, training with synthetic data does not consistently improve performance, and only few training-testing pairs showed increased accuracy. This shows that training with bootstrapped dataset has the potential to boost model performance during transferbility studies. 

\subsection{Symmetric and Asymmetric Transferability} \label{sec:sym-asym}

\begin{table}[t]
\centering
\captionsetup{justification=centering}
\caption{Observed Transferability}
\medskip
\label{tab:table-2}
   \begin{tabular}{|l|l|}
   \hline
   \textbf{Training Attack} &\textbf{Observed Transferability}
   \\ \hline
   DDoS (2)     &HULK (4)
   \\ \hline
   GoldenEye (3)      &DDoS(2), Hulk (4), \\
    & Slowhttptest (5), Slowloris (6)
   \\ \hline
    Hulk (4)      &DDoS (2)   
   \\ \hline
    Slowhttptest (5)      &GoldenEye (3), Slowloris (6)    
    \\ \hline
     Slowloris (6)      &GoldenEye (3), HULK (4),\\
     &Slowhttptest (5)  
    \\ \hline
    \end{tabular}
\end{table}

Here, we analyze the cases where the transferability relationships are symmetric and where they asymmetric. From Table \ref{tab:table-2}, we can observe the symmetric transferability relationships between Attacks 3, 5, and 6 reflect on the fact that Dos GoldenEye (Attack 3), DoS Slowloris (Attack 5) and DoS Slowhttptest (Attack 6) are very similar in the way they are performed. 

DoS Slowhttpattack and DoS Slowloris are essentially the same attack simulated using two different attack tools. In both, the attacker opens multiple HTTP connections with the target machine and keeps them open by declaring a large amount of data to send and then sending the data at a very slow rate. It keeps the connections to the target open and occupied making the target unavailable to legitimate requests. DoS GoldenEye is an attack tool that explores vulnerabilities in the target, it tests how susceptible a target is to a DoS attack. DoS GoldenEye opens multiple HTTP connections with the target server and keeps those connections open with keep alive headers and caching control options because of which the connections cannot be closed, making the server unavailable to other HTTP requests. Slow HTTP attacks also open multiple HTTP connections with the target and keep open connections by sending large volumes of data at a slow rate, thereby making the target unavailable to other users. Whether it is for the keep alive header or data being received slowly, the target is unable to close these connections. The same applies to Slowloris/Slowhttptest testing tools. This implies that although DoS GoldenEye is a flooding attack, it operates very similarly to slow HTTP attacks. This could explain why training with DoS GoldenEye transfers well for both Slow HTTP attacks.

Next, we observe that training our proposed model with DoS HULK (Attack 4) causes it to transfer for DDoS attacks (Attack 2) and vice versa. Training with DDoS attacks transfers for DoS HULK attack because the DDoS attack was generated with Low Orbit Ion Cannon that simulates UDP, TCP or HTTP flooding attacks by sending enormous traffic volume to the target machine. DoS HULK is a flooding attack where the target is flooded with HTTP data packets, sending GET requests and hitting its resource pool. Both these attacks are flooding attacks that may cause model learning to transfer across them..

\begin{table}[t]

    \centering
    \captionsetup{justification=centering}
    \caption{Number of Features Selected by RFE}
    \label{tab:table-3}
    \begin{tabular}{|l|c|c|c|c|c|c|}
    \hline
    \textbf{\begin{tabular}[c]{@{}l@{}}Attack \\ Pair\end{tabular}}      & (3,2) & (3,4) & (3,6) & (4,6) & (3,1) & (3,7) \\ \hline
    \textbf{\begin{tabular}[c]{@{}l@{}}Selected\\ Features\end{tabular}} & 4     & 6     & 9     & 14    & 21    & 24    \\ \hline
    \end{tabular}
\end{table}

On the other hand, some of the transferablity relationships are not symmetrical. Training the model with DoS GoldenEye (Attack 3) causes the model to transfer for DDoS (Attack 2) and DoS HULK (Attack 4), however this observed correlation does not apply the other way around. Similarly, training the model with DoS Slowloris (Attack 6) causes it to transfer for DoS HULK (Attack 4) but not vice versa.To explain this, we hypothesize that correlated attacks will have few dominant features selected by the feature selection algorithm which may explain the correlations between the attacks whereas the uncorrelated attacks will have many features selected by the RFE algorithm indicating that there are not many dominant features that are important for learning which could explain the low correlations. This can be strengthened by the results in Table \ref{tab:table-3} for attack pairs (3,2), (3,4), and (3,6) that have 4, 6 and 9 features selected by the algorithm. Each of these attack pairs are correlated. However for uncorrelated attacks like (3,7) and (3,1), the RFE algorithm selects 24 and 21 features which are considerably more.

\subsection{Feature Selection}

Here, we analyze some of the features selected by the RFE algorithm for the training-testing attack pairs shown in Table \ref{tab:table-3}. Some of the features selected by the algorithm for the training-testing attack pairs are \textit{Destination Port}, \textit{Backward Packet Length Std}, \textit{Flow IAT Mean}, \textit{Subflow Backward Bytes}, \textit{Average packet size} and \textit{Packet Length mean}. \textit{Destination Port} is the only feature that was selected by all training-testing attack pairs. Attacks targeting a specific service are often targeted to a particular port that runs the service. For example: an attack that leverages the File Transfer Protocol could use ports 20 and 21. Similarly, often web attacks will have attack data packets flooding ports 8080 or 8843 of the victim server.

\subsection{Effect of Preprocessing}

\begin{figure}[t]
    \centering
    \includegraphics[width=0.48\textwidth]{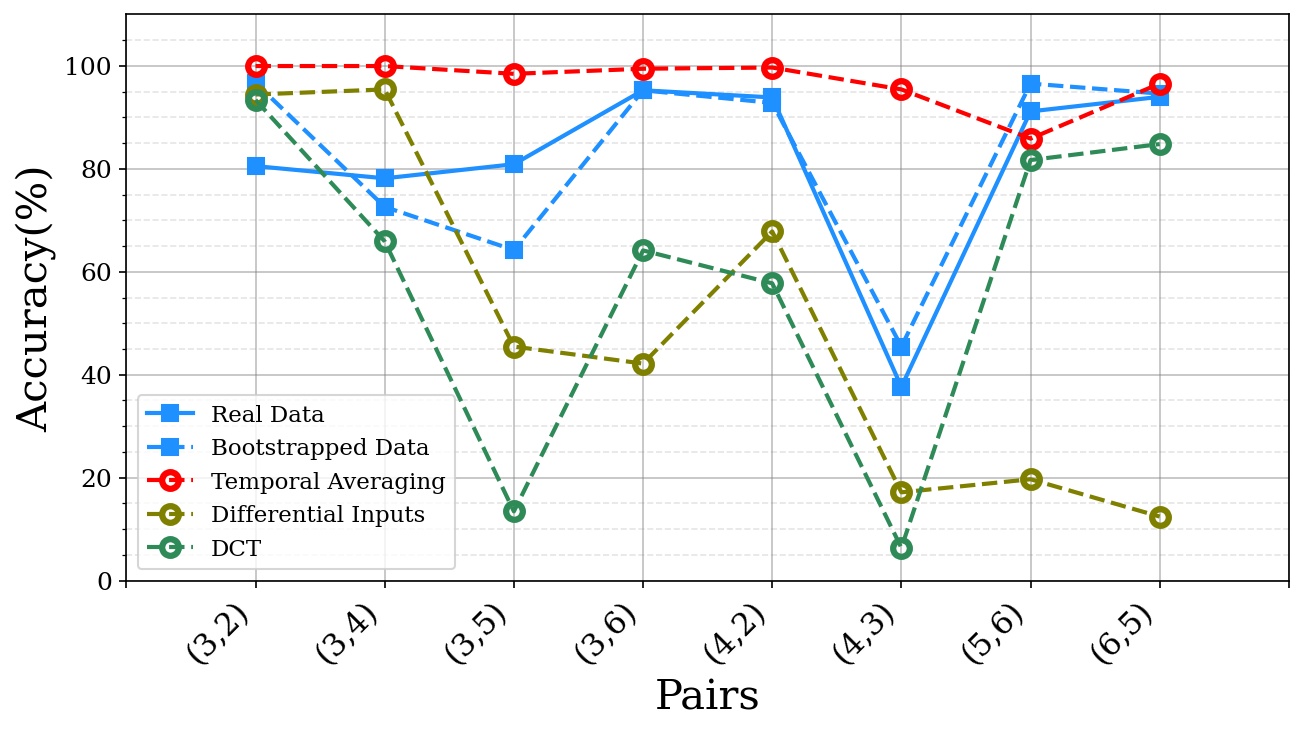}
    \caption{Attack Accuracy comparison on selected training and testing pairs for different preprocessing approaches }
    
    \label{fig:preprocessing}
\end{figure}

Finally, we investigate the effects of data preprocessing on transferability relationships. We can see from Fig. \ref{fig:preprocessing} that the model exhibits some of the same transferability relationships as listed in Table \ref{tab:table-2} when using data preprocessing. Temporal averaging boosts the transferability accuracy of the shown training-testing attack pairs over differential inputs, DCT, real data and bootstrapped data. Therefore, we identify temporal averaging as a potential data preprocessing technique to boost the accuracy of attack detections.

\section{Conclusion}

In this work, we develop a deep neural network that classifies attack data packets in the CICIDS-2017 dataset with high accuracy. We use this architecture to study transferability in learning and analyze the symmetric or asymmetric nature of transferability relationships. We utilize the recursive feature selection based analysis to explain the transferability relationships and analyse some selected features. Further, we provide hypotheses for observing correlations in transferablility relationships that we validate with our findings. We examine the effects of bootstrapping and training with synthetically generated data. We also study different preprocessing techniques and identify one technique that boosts model performance for certain training-testing pairs. This work can be further expanded to observing transferability in different settings, for example using other datasets, or a different deep learning paradigm (for example, in distributed learning).

Studies could also focus on analyzing trends in specific or selected features in studying transferability correlations. It is interesting to observe that we only found correlations in attacks that are different types of DoS attacks, however the same could not be observed for other kinds of attacks in the dataset such as Botnets, FTP-Patator, Heartbleed, Infiltration and SQL Injection attacks. Future work can focus on identifying attributes that could boost model learning and enable model learning to be transferable to these kinds of attacks.

\ifCLASSOPTIONcaptionsoff
  \newpage
\fi
\bibliographystyle{IEEEtran} 

\bibliography{nattack2021}

\end{document}